\documentclass[a4paper]{jpconf}
\usepackage{graphicx}
\def\be{\begin{equation}}
\def\ee{\end{equation}}
\def\beq{\begin{equation}}
\def\eeq{\end{equation}}
\def\bc{\begin{center}}
\def\ec{\end{center}}
\def\bea{\begin{eqnarray}}
\def\eea{\end{eqnarray}}

\begin{document}
\hfill{\em dedicated to Guido Altarelli}\\
\vskip 0.5 cm

\title{Aspects of Leptonic Flavour Mixing}

\author{Ferruccio Feruglio}

\address{Dipartimento di Fisica e Astronomia `G.~Galilei', Universit\`a di Padova
\\
INFN, Sezione di Padova, Via Marzolo~8, I-35131 Padua, Italy}

\ead{feruglio@pd.infn.it}

\begin{abstract}
Since the discovery of neutrino oscillations many ideas have been put forward  to explain the special features of the leptonic mixing and the differences with respect to the quark sector.
In this talk I review some of these proposals, emphasizing especially their predictability. In the light of the new data, I first revisit fixed-point relations among mixing angles and phases. Then
I briefly comment on radiative neutrino masses. Finally I discuss the role of flavour symmetries. Given the very many existing models I focus on two classes of models. On the one hand 
I illustrate the ability of models based on a generalization of the anarchy idea in reproducing the main features of both the quark and the lepton spectrum, also in a GUT framework.
On the other hand I discuss less ambitious but more predictive models based on discrete flavour symmetries, centered on the properties of the leptonix mixing matrix.
\end{abstract}
\section{Introduction}
The explanation of the leptonic flavour mixing is an aspect of a more general problem, the so called ``flavour puzzle'', the lack of understanding of the variety
of Yukawa couplings needed to accommodate fermion masses and mixing angles in the Standard Model (SM). Mass ratios of charged fermions have unexplained hierarchies.
Neutrinos masses are extremely small compared to the charged fermion ones and lepton mixing angles have apparently no relationship to the quark mixing angles,
despite the fact that in grand unified theories (GUT), where fermion quantum numbers find a natural justification, there is no fundamental distinction
between leptons and quarks. The neutrino sector is very special since it is the only one where predictions are still possible. Among the open questions we have the nature - Dirac versus Majorana -  of neutrinos, the mass ordering, the absolute mass scale, the CP violating phases.
Given the success of the SM in describing accurately strong and electroweak interactions, the answers to these questions have long been pursued within the framework of a 
quantum field theory, extending and completing the SM. In such a context, combinations of masses and mixing angles can be computed in terms of other non-vanishing input parameters
in a small number of cases. The most relevant ones are:
\begin{itemize}
\item[i)] {\bf Fixed-point relations}\hfill\break\noindent 
In this case the mass parameters of the theory are unknown and originate at a very high energy scale. At the same time the renormalisation group flow drives these input parameters to special low-energy values, which do not depend on the initial conditions at the high scale.
We speak of an infrared stable fixed point of the renormalization group equations (RGE).
\item[ii)] {\bf Radiative fermion masses}\hfill\break\noindent
A combination of mass parameters vanishes accidentally at the classical level as a consequence of the specific particle content and renormalizability. 
Unless protected by a symmetry of the theory, such a combination will in general get a non-vanishing contribution from quantum corrections which,
in the small coupling regime, can be computed in perturbation theory.
\item[iii)] {\bf Symmetries}\hfill\break\noindent 
There are cases in which relations among mass parameters are implied by a symmetry. The ideal case is represented by exact symmetries, such as the abelian gauge invariance in quantum electrodynamics, requiring 
a vanishing photon mass. Unfortunately exact symmetries do not apply to fermion masses and mixing angles. For instance, the largest non-abelian global symmetry of the quark sector, $SU(3)^3$, is completely broken
by the SM Yukawa couplings. Not even in the lepton sector we have hints for exact symmetries. We are thus lead to consider approximate symmetries, like for instance isospin which implies the near
equality between the proton and the neutron mass. This results in a huge number of possibilities where symmetries can be abelian or non-abelian, continuous or discrete, global or local, explicitly or spontaneously broken.
\end{itemize}
There can also be an interplay among these possibilities. In the following I will comment the three cases with emphasis on the lepton mixing parameters.

\section{Lepton mixing from RGE flow}
\begin{figure}[h]
\begin{minipage}{16pc}
\includegraphics[width=16pc]{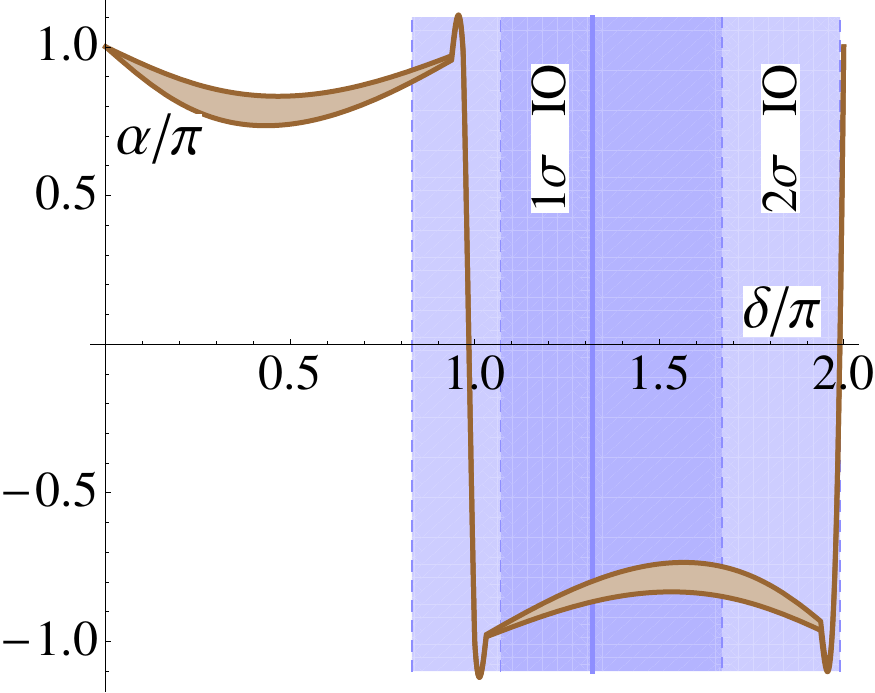}
\caption{\label{label1} Region allowed by the fixed point relation (6) in the plane $(\delta/\pi,\alpha/\pi)$ (brown). In violet the values of $\delta$ preferred by a global fit to neutrino oscillations (pre-Neutrino 2016).}
\end{minipage}\hspace{4pc}%
\begin{minipage}{18pc}
\includegraphics[width=18pc]{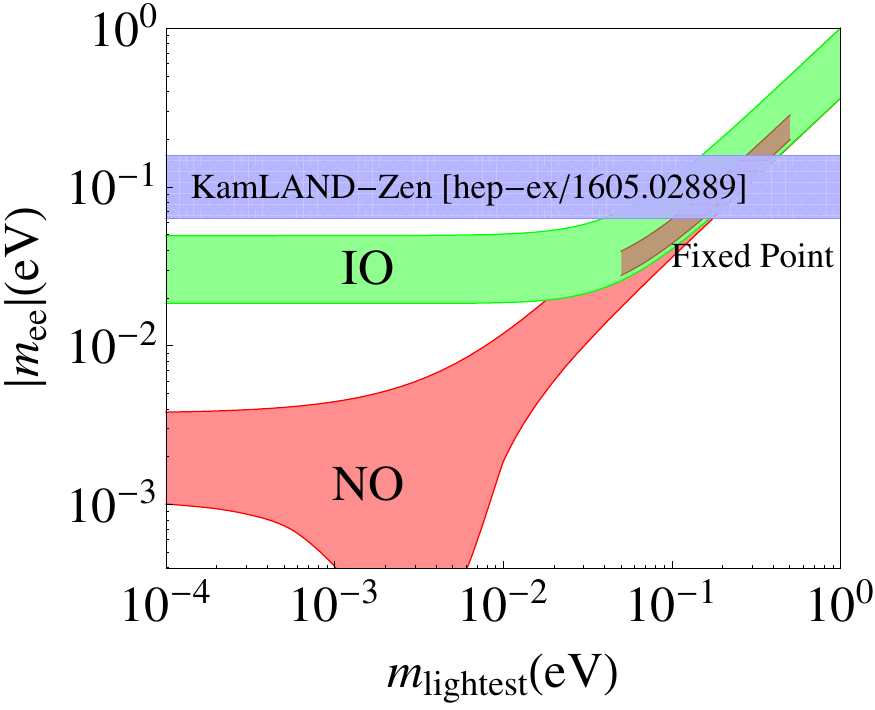}
\caption{\label{label2} Prediction of $|m_{ee}|$ for the case of inverted ordering (brown), when the fixed point relation (6) is verified.}
\end{minipage} 
\end{figure}

Suppose that neutrino masses are of Majorana type and come from the Weinberg operator 
\begin{equation}
{\cal L}_5=\frac{1}{\Lambda}(\varphi^\dagger l)^Tw(\varphi^\dagger l)~~~,
\label{weinberg}
\end{equation}
originating at some energy scale $\Lambda$ much larger than the electroweak scale $v$. This can occur, for instance, via the see-saw mechanism. At low energies neutrino masses are of order $v^2/\Lambda$, specified by
the parameters of the symmetric matrix $w$. The elements of the mixing matrix $U_{PMNS}$ vary along the RGE trajectories with a speed controlled by
the combination:
\begin{equation}
\frac{\eta}{16\pi^2}y_\tau^2~\frac{m_i+m_j}{m_i-m_j}~\approx~ 5\times 10^{-7}~\tan^2\beta~\frac{m_i+m_j}{m_i-m_j}~~~,
\end{equation}
where $m_i$ are the neutrino masses, $y_\tau$ is the tau Yukawa coupling and $\eta$ a numerical factor. ($\eta=-3/2$, $y_\tau=\sqrt{2} m_\tau/v$ in the SM, $\eta=1$, $y_\tau=\sqrt{2} m_\tau/(v \cos\beta)$ in the MSSM). Fixed points are reached only if this speed is sufficiently large, which in turn requires a strong degeneracy among neutrino masses. Fixed point relations for the elements of the lepton
mixing matrix have been studied long ago \cite{Chankowski:2001mx} and in the CP-conserving case they are summarized by the following equality among the lepton mixing angles $\theta_{ij}$:
\begin{equation}
\sin^2 2\theta_{12}=\sin^2\theta_{13}\frac{\sin^2 2\theta_{23}}{(\sin^2\theta_{13}+\sin^2\theta_{23}\cos^2\theta_{13})^2}~~~.
\label{CPconserving}
\end{equation}
Data rule out this relation by many standard deviations. The allowed 3$\sigma$ ranges for the left-hand and right-hand sides are ($0.75\div 0.92$) and ($0.05\div 0.16$), respectively. In the CP-violating regime the only acceptable case is when $m_1$ and $m_2$ are
nearly degenerate and the corresponding fixed-point relation is \cite{Casas:1999tg}
\begin{equation}
Re(U^*_{31} U_{32})=0~~~,
\label{cpv}
\end{equation}
which gives back eq. (\ref{CPconserving}) when phases are neglected. It is useful to translate eq. (\ref{cpv}) into a relation satisfied by
phases and mixing angles. 
Here I adopt the convention for Majorana phases that allows to eliminate the Dirac phase $\delta$ from the parameter $|m_{ee}|$ relevant to neutrino-less double beta decay:
\begin{equation}
|m_{ee}|=|c^2_{12}c^2_{13} m_1+s^2_{12}c^2_{13} e^{i \alpha}m_2+s^2_{13}e^{i\beta}m_3|~~~.
\end{equation}
I find that eq. (\ref{cpv}) explicitly reads:
\begin{equation}
s_{12}c_{12}(c^2_{23}s^2_{13}-s^2_{23})\cos\alpha/2+s_{13}s_{23}c_{23}\left[c^2_{12}\cos(\alpha/2-\delta)-s^2_{12}\cos(\alpha/2+\delta)\right]=0 ~~~,
\label{CPviolating}
\end{equation}
where $s_{12}=\sin\theta_{12},...$ Given our good knowledge of the mixing angles, the relation (\ref{CPviolating}) translates into a constraint on the Dirac phase $\delta$ and the Majorana phase $\alpha$, which I show in fig. \ref{label1}.
The preferred value of $\alpha$ is around $-150^0$. This in turn has an impact on neutrino-less double beta decay. If the ordering is inverted, which is the most plausible possibility
if $m_1$ and $m_2$ are strongly degenerate at the scale $\Lambda$, the mass parameter $|m_{ee}|$ is predicted to lie in a small region close to the present upper limit, as shown in fig. \ref{label2}.
This mechanism cannot be used to explain the smallness of $\theta_{13}$ or the largeness of $\theta_{23}$ since these two angles do not appreciably run in the regime $m_1\approx m_2$, but the correlations implied by the above scenario (large scale $\Lambda$, inverted hierarchy, $\pi\le\delta\le 2\pi$ related to $\alpha\approx -150^0$, sum of neutrino masses
of order 0.1 eV and $|m_{ee}|$ close to the present bounds) are certainly interesting.
\section{Radiative Neutrino Masses}
\begin{figure}[h]
\begin{minipage}{16pc}
\includegraphics[width=16pc]{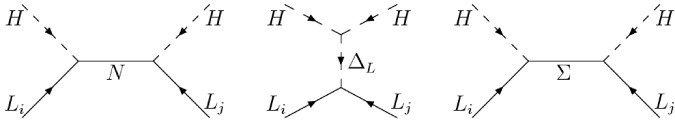}
\caption{\label{label3} The three types of see-saw mechanism.}
\end{minipage}\hspace{4pc}%
\begin{minipage}{18pc}
\includegraphics[width=16pc]{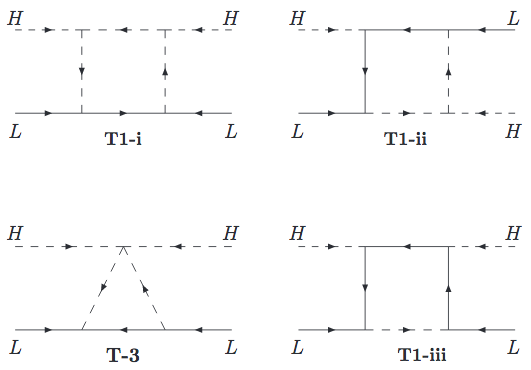}
\caption{\label{label4} Independent diagrams giving rise to the Weinberg operator, eq. (1), at 1-loop, from ref.  \cite{Bonnet:2012kz}.}
\end{minipage} 
\end{figure}

The simplest origin of the Weinberg operator in eq. (\ref{weinberg}) is the see-saw mechanism,
i.e. the tree-level exchange of an heavy multiplet between Higgs and lepton doublets. The possibilities are exhausted by the tree types of see-saw, depending on the quantum numbers of the heavy mediator:
a fermion singlet, a scalar triplet or a fermion triplet, see fig. \ref{label3}. These new states are easily embedded into multiplets of GUTs, where the see-saw mechanism finds its more natural realization.
In a special class of models, the specific particle content can forbid the see-saw mechanism, while allowing the Weinberg operator to arise at $L\ge 1$ loop order. Neutrinos are massless at the classical level and pick up their masses
from quantum corrections. The topologies of the diagrams contributing to the Weinberg operator have been classified up to 2-loop order \cite{Babu:2001ex,Bonnet:2012kz,Angel:2012ug,Sierra:2014rxa}. 
The independent one-loop diagrams  \cite{Bonnet:2012kz} are displayed in fig. \ref{label4}.
At least two new multiplets are required as intermediate states \cite{Law:2013dya}.
In this scenario neutrino masses are suppressed by a loop factor, $(1/16\pi^2)^L$, and the intermediate states running in the loop can be sufficiently light to be probed at existing facilities,
at variance with the typically heavy states of the see-saw mechanism. This is the main motivation of the framework. Neutrino physics can become directly accessible at high-energy colliders,
with the production of new particles with masses in the TeV range. The new states are also responsible for lepton flavour violation (LFV) both at the tree-level (e.g. $\mu\to 3e$)
and at one-loop (e.g. $\mu\to e \gamma$), that can be searched for at present or future high-intensity facilities. In the presence of suitable symmetries some of the intermediate states can also provide a dark matter candidate \cite{Farzan:2012ev}.
The related phenomenology is extremely interesting and so wide that cannot be reviewed here \cite{Boucenna:2014zba}.
Such a rich benefit does not come without a cost, at least from the viewpoint of the flavour problem.
As soon as we go beyond the tree level we have a large variety of possible realisation of the Weinberg operator at the microscopic level. There are many independent loop diagrams involved: four at 1-loop,
twenty at 2-loop and so on. For each given diagram there are several different choices of intermediate states. The number of microscopic models rapidly diverges and the uniqueness of the tree-level see-saw is lost.
Also the number of independent parameters characterising masses and interactions of the new states increases. Moreover these parameters must be tuned both to reproduce neutrino masses and mixing angles and
to cope with the present bounds on LFV. The flavour problem gets amplified. Finally, relatively light intermediate states makes gauge coupling unification more difficult, the successful examples requiring a rather ad-hoc particle content \cite{Hagedorn:2016dze}.
\section{A minimum amount of flavour symmetry}
As a matter of fact, we have no evidence for striking hierarchies among lepton mixing angles or neutrino masses. This led to the idea of ``anarchy'' \cite{Hall:1999sn,Haba:2000be,deGouvea:2003xe,Espinosa:2003qz,deGouvea:2012ac,Brdar:2015jwo} which, at the level
of the neutrino mass matrix, can be roughly formulated by requiring that all matrix elements are of the same order, with no particular pattern. To some extent 
this idea is consistent with data and it suggested that the angle $\theta_{13}$ had to be close to the present value, well before its measurement. Should the atmospheric
angle $\theta_{23}$ deviate from maximal, as indicated -- even though not conclusively -- by the most recent data, this scenario would be further reinforced.
It is natural to ask if we can adopt the same principle also for quarks and charged leptons and start from a theory where the Yukawa couplings are described by anarchical
$3\times 3$ matrices with order one matrix elements. Can the approximate regularities of the charged fermion sector emerge from this initial chaos?

It is interesting to analyze this question in the context of GUTs where lepton and quarks are closely related. GUTs have many good properties.
In GUTs particle classification is greatly clarified. Quarks and leptons of the same generation belong to few multiplets of the grand unified group,
a single representation being sufficient in the case of SO(10). Charge quantisation and gauge anomaly cancelation, which look miraculous within the SM, are neatly explained. 
Here we focus on an SU(5) GUT. In a minimal formulation of this theory,
matter fields are described by three copies of the $10=(q,u^c,e^c)$ and $\bar 5=(l,d^c)$ representations, while the Higgs fields $\varphi$ and $\bar \varphi$ transform as $5$ and $\bar 5$, respectively.
Fermion masses are described by the Yukawa interactions:
\begin{equation}
{\cal L}_Y=10~ y_u~ 10~ \varphi+\bar 5~ y_d~ 10~ \bar\varphi+\frac{1}{M}\bar 5~ w~ \bar 5~ \varphi \varphi+...
\end{equation}
where $y_{u,d}$ and $w$ are matrices in generation space and $M$ is a large scale, possibly close to the GUT scale. After electroweak symmetry breaking the first term describes up-quark masses, the last one is the grand unified version of the Weinberg operator in eq. (\ref{weinberg}). The second term describes at the same time down-quark masses and charged lepton masses, which are equal at the GUT scale in this approximations. Corrections
to this relation are provided by additional contributions denoted by dots. The anarchy principle translates into the requirement that
the matrices $y_{u,d}$ and $w$ have entries of the same order of magnitude, with no built-in structure.

An appealing mechanism by which the hierarchy observed in the charged fermion sector can be produced,
starting from anarchical matrices $y_{u,d}$, is a rescaling of the matter fields:
\begin{equation}
10\to F_{10}~ 10~~~,~~~~~~~~~~~~~~~\bar 5\to F_{\bar 5}~ \bar 5~~~.
\end{equation}
Here $F_{10,\bar 5}$ are diagonal matrices of the type
\begin{equation}
F_X=
\left(
\begin{array}{ccc}
\epsilon_X'&0&0\\
0&\epsilon_X&0\\
0&0&1
\end{array}
\right)~~~~~~~~~~(1\ge \epsilon_X\ge \epsilon_X')~~~.
\end{equation}
For instance, after rescaling the 10 representations, the effective matrix of Yukawa couplings for the up quarks becomes
\begin{equation}
{\cal Y}_u=F_{10}~ y_u~ F_{10}~~~,
\end{equation}
which is hierarchical and nearly diagonal if $1\gg \epsilon_{10}\gg \epsilon_{10}'$. By adjusting the suppression factors $\epsilon_{10}$ and $\epsilon_{10}'$ we can match the up-quark masses
and generate small contributions to the quark mixing angles.  Such a mechanism is rather generic in model building. The rescaling matrices $F_{X}$ can arise
in a variety of frameworks such as models with an abelian flavour symmetry, models with an extra dimension and
models with partial compositeness or specific conformal dynamics \cite{Feruglio:2015jfa}. 

Since the mass hierarchy in the down-quark and charged-lepton sectors is much less pronounced than in the 
up-quark sector, we need a milder rescaling from $F_{\bar 5}$. As a useful reference we can choose
\begin{equation}
F_{\bar 5}=\left(
\begin{array}{ccc}
1&0&0\\
0&1&0\\
0&0&1
\end{array}
\right)~~~.
\label{Fan}
\end{equation}
In this limit we find
\begin{equation}
m_u:m_c:m_t\approx m_d^2:m_s^2:m_b^2\approx m_e^2:m_\mu^2:m_\tau^2
\end{equation}
which is approximately correct, at least at the level of orders of magnitude. Moreover, at the leading order we have
\begin{equation}
{\cal Y}_e={\cal Y}_d^T~~~,
\label{tran}
\end{equation}
where both ${\cal Y}_{e}$ and ${\cal Y}_{d}$ are lopsided matrices since $F_{\bar 5}\ne F_{10}$. The relation (\ref{tran}) should be corrected since it leads to wrong mass equalities
for the first two generations. The required corrections are sizeable, but not huge and (\ref{tran}) can still be valid at the level of orders of magnitude.
In the limit where (\ref{tran}) is exact, it predicts a small contribution to the quark left-handed mixing and a large contribution to the lepton left-handed mixing, which is exactly
what we observe. For the right-handed components a large (small) mixing for quarks (leptons) is predicted, which however is not observable at low energies.
 
The neutrino mass matrix is $m_\nu\propto F_{\bar 5}w F_{\bar 5}~ v^2/M$. When (\ref{Fan}) holds neutrino mass ratios and mixing angles 
reproduce exactly the case of anarchy, since they are generated from the random, order-one, matrix elements of $w$.
However within the extreme choice in eq. (\ref{Fan}) there is no preference for the type of neutrino mass ordering and 
no explanation of the smallness of $\sin^2\theta_{13}$ and $\Delta m^2_{sol}/\Delta m^2_{atm}$. It is worth to replace (\ref{Fan}) 
by a more generic possibility, such as
\begin{equation}
F_{\bar 5}=\left(
\begin{array}{ccc}
\lambda^{Q_1}&0&0\\
0&\lambda^{Q_2}&0\\
0&0&1
\end{array}
\right)~~~.
\end{equation}
Here $\lambda$ is an expansion parameter, typically smaller than 0.5 and $Q_{1,2}$ are two positive charges, $Q_{1}\ge Q_{2}\ge 0$. Anarchy is reproduced
when $Q_{1,2}=0$. It is not surprising that several examples with $Q_{1}$ non vanishing can be found where a small $\theta_{13}$ is more easily reproduced than in anarchy \cite{u1,u2,u3,u4,u5}, see fig. 5. In all the more successful
examples the normal ordering of neutrino masses is preferred. First hints of such a preference are currently shown in global fits to neutrino oscillation experiments  \cite{Capozzi:2016rtj}.
\begin{figure}[h]
\begin{minipage}{16pc}
\includegraphics[width=18pc]{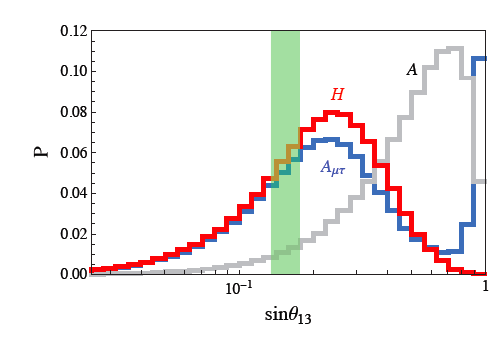}
\caption{\label{label5} Probability distribution of $\sin\theta_{13}$, for several choices of $F_{\bar 5}$, from ref. \cite{u3}:  Anarchy [$A$, $(\lambda,Q_1,Q_2)=(0.2,0,0)$],
$\mu\tau$-Anarchy [$A_{\mu\tau}$, $(\lambda,Q_1,Q_2)=(0.2,1,0)$], Hierarchy [$H$, $(\lambda,Q_1,Q_2)=(0.4,2,1)$].}
\end{minipage}\hspace{4pc}%
\begin{minipage}{18pc}
\includegraphics[width=18pc]{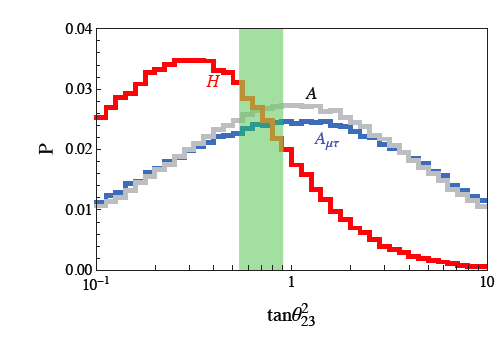}
\caption{\label{label6} Probability distribution of $\tan^2\theta_{23}$, for several choices of $F_{\bar 5}$, from ref. \cite{u3}:  Anarchy [$A$, $(\lambda,Q_1,Q_2)=(0.2,0,0)$],
$\mu\tau$-Anarchy [$A_{\mu\tau}$, $(\lambda,Q_1,Q_2)=(0.2,1,0)$], Hierarchy [$H$, $(\lambda,Q_1,Q_2)=(0.4,2,1)$].}
\end{minipage} 
\end{figure}

These results are impressive. 
All qualitative features of quarks and lepton masses and mixing angles are reproduced. The difference between the two mixing matrices, $V_{CKM}$ and $U_{PMNS}$, is neatly explained.
The amount of symmetry required is minimal. Apart from the GUT symmetry that connects members of the same fermion generation,
the observed intergenerational hierarchies are all generated by few rescaling factors, which could arise even without an underlying flavour symmetry. 
The same mechanism works for SO(10) GUTs as well \cite{Kitano:2003cn,Feruglio:2014jla}.
By extending the model by the inclusion of a set of right-handed neutrinos, leptogenesis successfully occurs \cite{Lu:2014cla}.
Anarchy arises as a special case. Though rather appealing at first sight, this approach has clear limitations. 
The most severe one is that the entries of the matrices $y_{u,d}$ and $w$ are independent order-one parameters.
Predictions for the various physical quantities can only be formulated in terms of broad distributions, assuming some statistical distribution for the unknown matrix elements of $y_{u,d}$ and $w$, see figs. 5 and 6.
Models in this class typically predict nearly flat distributions for the CP violating phases. Thus features such as
the closeness of the Dirac CP phase to the maximal value are purely accidental in this framework.
It is not possible to go beyond order-of magnitude estimates, whereas today we have precise data and we would like to have models whose predictions can be tested
at the level of accuracy reached by the present experiments. 
\section{More symmetry?}
More predictive frameworks typically require more symmetries. Model building has been largely influenced by features such as
the smallness of $\theta_{13}$, the closeness of the atmospheric angle to the maximal value and, more recently, the indication of a maximal Dirac CP phase.
Several forms of quark-lepton complementarity have also been invoked \cite{Minakata:2004xt,Antusch:2005ca,Mohapatra:2006gs}. If some of these features are not accidental, they can guide us in the search for a fundamental
principle governing the flavour sector. Several symmetric patterns of lepton mixing angles have been suggested in the past, such as the tribimaximal (TB) mixing
or the bimaximal (BM) mixing:
\begin{equation}
U_{TB}=\left(
\begin{array}{ccc}
\sqrt{\frac{2}{6}}&\frac{1}{\sqrt{3}}&0\\
-\frac{1}{\sqrt{6}}&\frac{1}{\sqrt{3}}&\frac{1}{\sqrt{2}}\\
-\frac{1}{\sqrt{6}}&\frac{1}{\sqrt{3}}&-\frac{1}{\sqrt{2}}
\end{array}
\right)~~~,~~~~~~~
U_{BM}=\left(
\begin{array}{ccc}
\frac{1}{\sqrt{2}}&\frac{1}{\sqrt{2}}&0\\
-\frac{1}{2}&\frac{1}{2}&\frac{1}{\sqrt{2}}\\
\frac{1}{2}&-\frac{1}{2}&\frac{1}{\sqrt{2}}
\end{array}
\right)~~~.
\end{equation}
They incorporate some of the above-mentioned aspects. These patterns can be adopted as first order approximations to the true lepton mixing matrix $U_{PMNS}$.
In this approach $U_{PMNS}$ is  expanded around a leading order matrix $U^0_{PMNS}$, which can coincide with $U_{TB}$, $U_{BM}$ or some other
symmetrical form:
\begin{equation}
U_{PMNS}=U^0_{PMNS}+...
\end{equation}
where dots stand for corrections. It is not difficult to identify flavour symmetries leading to $U^0_{PMNS}$.
For example discrete flavour symmetries showed very efficient in reproducing
$U_{TB}$, $U_{BM}$ or other leading order patterns. These constructions require small non-abelian permutation groups, such as $A_4$ and $S_4$.
In the so-called direct approach we can predict the three mixing angles and the CP violating phase, while neutrino masses
are only constrained within extended ranges and are fitted by adjusting the free parameters \cite{Altarelli:2005yp,Altarelli:2005yx,Altarelli:2010gt,King:2013eh}. 
Discrete flavour symmetries are also relevant in the so called indirect models \cite{King:2013eh}. In this case the breaking of the flavour group leaves no residual
symmetries and its role is mainly to get specific vacuum alignments of the scalar fields that control fermion masses. 

Today we know that the leading order patterns require sizable corrections. This may come from an additional rotation in generation space. For instance
we can perturb the bimaximal mixing $U_{BM}$ by a rotation $U_{12}$ among the first two generations, coming from the diagonalization of the charged lepton sector:
\be
U_{PMNS}=U_{12}(\alpha,\delta) U_{BM}=
\left(
\begin{array}{ccc}
\cos\alpha&e^{-i\delta}\sin\alpha&0\\
-e^{i\delta}\sin\alpha&\cos\alpha&0\\
0&0&1
\end{array}
\right) U_{BM}~~~.
\label{bimaxmod}
\ee 
The mixing angles and the Dirac phase are predicted in terms of $(\alpha,\delta)$ and we get two relations among
physical quantities, known as sum rules:
\bea
\sin^2\theta_{12}&=&\frac{1}{2}+\sin\theta_{13}\cos\delta_{CP}+O(\sin^2\theta_{13})\label{s12}\\
\sin^2\theta_{23}&=&\frac{1}{2}+O(\sin^2\theta_{13})
\eea
\begin{figure}[h]
\begin{minipage}{16pc}
\includegraphics[width=16pc]{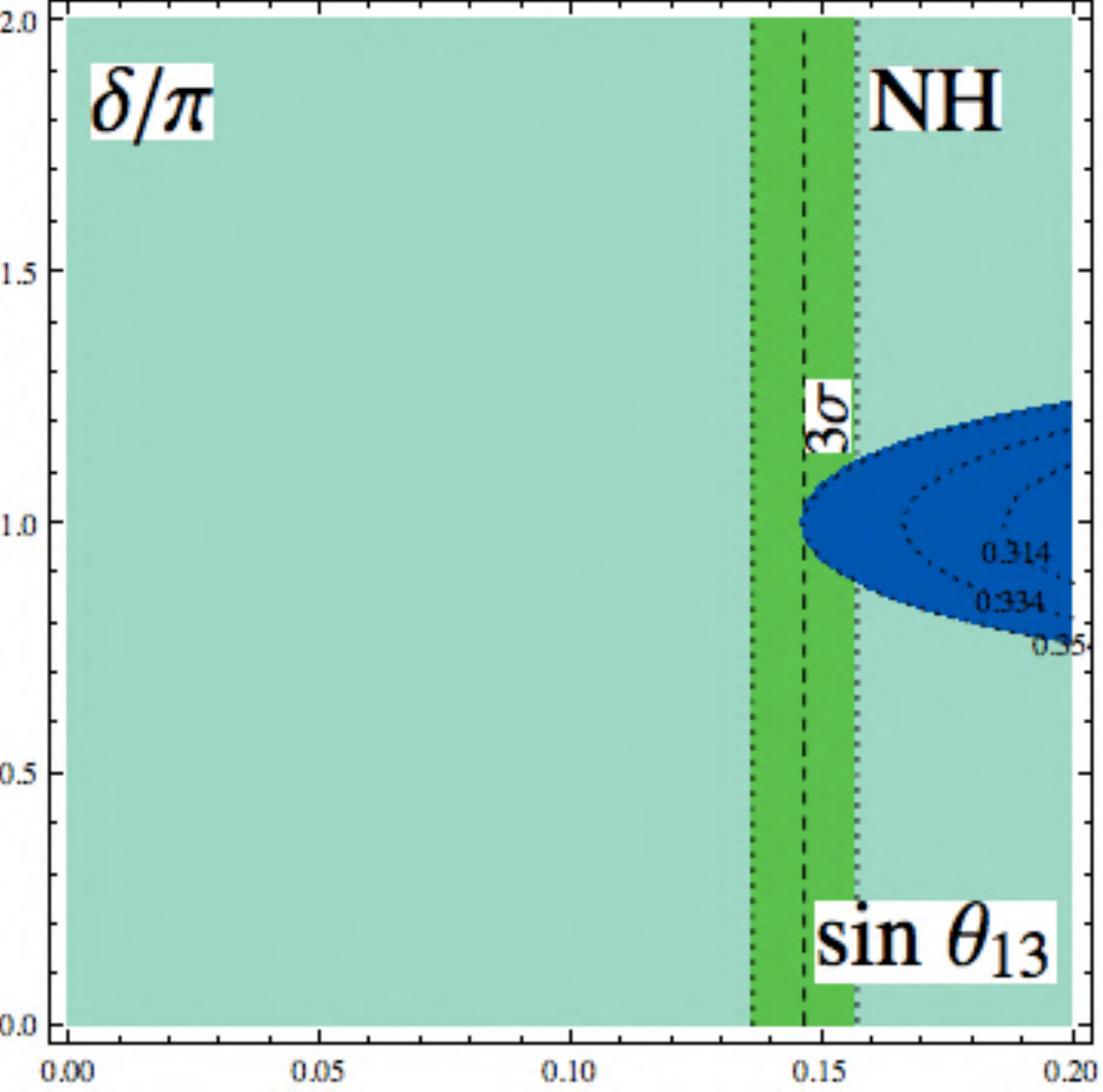}
\caption{\label{label7} Contours of $\sin^2\theta_{12}$ in the plane $(\sin\theta_{13},\delta/\pi)$ from eq. (\ref{s12}). The blue region is allowed at 3$\sigma$ by the measurement of $\sin^2\theta_{12}$.}
\end{minipage}\hspace{4pc}%
\begin{minipage}{16pc}
\includegraphics[width=16pc]{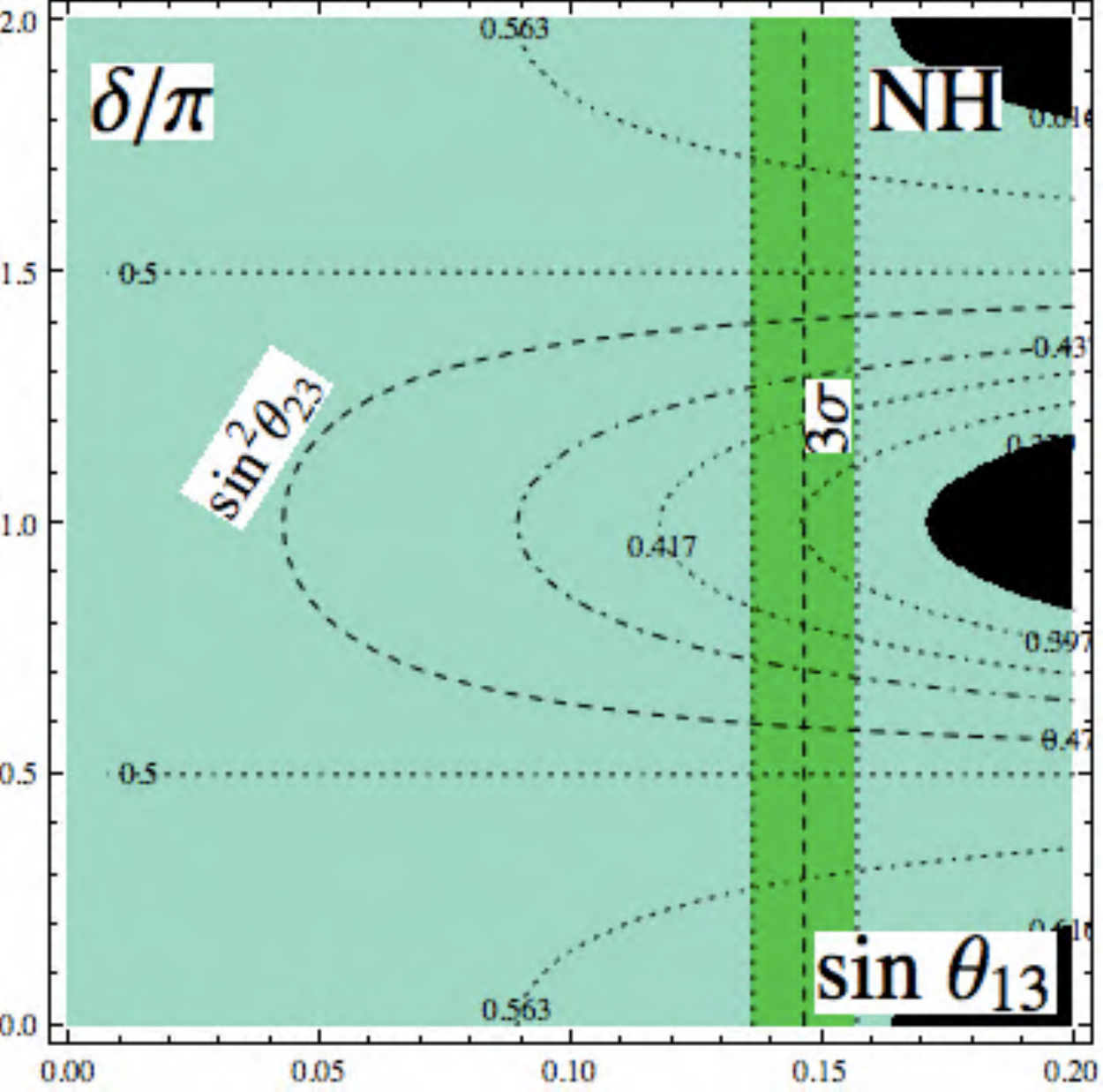}
\caption{\label{label8} Contours of $\sin^2\theta_{23}$ in the plane $(\sin\theta_{13},\delta/\pi)$ for the sum rule arising from $TM_2$. The black region is ruled out at 3$\sigma$.}
\end{minipage} 
\end{figure}

\noindent
This framework predicts $\theta_{23}$ close to maximal and $\delta_{CP}$ close to $\pi$ in order to reproduce correctly  $\sin^2\theta_{12}$, as can be seen from fig. (\ref{label7}).
Another possibility consists in modifying the TB mixing by rotations that give rise to a non-vanishing $\theta_{13}$:
\be
U_{TM_2}=U_{TB}~U_{13}(\alpha,\delta)~~~~~~~~~~~~~~~~~~~U_{TM_1}=U_{TB}~U_{23}(\alpha,\delta)~~~,
\label{trimax}
\ee 
where $U_{13}(\alpha,\delta)$ and $U_{23}(\alpha,\delta)$ are the transformations analogous to $U_{12}(\alpha,\delta)$, acting in the 13 and 23 planes, respectively. These
mixing pattern are called trimaximal.
The corresponding sum rules are shown in Table \ref{tt3} \cite{Ballett:2013wya}.
The interesting feature of these relations is that the predicted deviations from TB are linear in $\sin\theta_{13}$ for $\sin^2\theta_{23}$, and quadratic for $\sin^2\theta_{12}$, 
known with much better precision. One of these relations is plotted in fig. \ref{label8} in the case of $TM_2$, from which we see that a substantial improvement in the data is needed
to test this possibility.
\begin{table}[h!]
\caption{\label{tt3} Sum rules for $TM_{1,2}$ mixing patterns.}
\begin{center}
\begin{tabular}{ll}
\br
$TM_1$& $TM_2$\\
\mr
$\sin^2\theta_{12}=\frac{1}{3}-\frac{2}{3}\sin^2\theta_{13}+O(\sin^4\theta_{13})$&$\sin^2\theta_{12}=\frac{1}{3}+\frac{1}{3}\sin^2\theta_{13}+O(\sin^4\theta_{13})$\\
$\sin^2\theta_{23}=\frac{1}{2}-\sqrt{2} \sin\theta_{13} \cos\delta_{CP}+O(\sin^2\theta_{13})$&$\sin^2\theta_{23}=\frac{1}{2}+\frac{1}{\sqrt{2}} \sin\theta_{13} \cos\delta_{CP}+O(\sin^2\theta_{13})$\\
\br
\end{tabular}
\end{center}
\end{table}

Instead of adding corrections to $U_{BM}$ or $U_{TB}$, we can look for flavour groups giving rise to
a mixing matrix $U^0_{PMNS}$ closer to the present data than $U_{BM}$ or $U_{TB}$.
Very remarkably, a complete classification of all possible mixing matrices $|U^0_{PMNS}|$ generated from any finite group has been recently carried out in ref. \cite{Fonseca:2014koa}.
Mixing angles close to the observed ones can be obtained by appealing to sufficiently large groups  (e.g. one group of the series $\Delta(6 n^2)$) and the corresponding patterns are of trimaximal type. 
In such cases the Dirac CP phase is trivial, which is disfavored by the present data. 

Another development consists in combining discrete and $CP$ symmetries \cite{Feruglio:2012cw,Holthausen:2012dk,Hagedorn2016} and exploring the symmetry breaking patterns such a combination can give rise to. 
A well-known example is that of the so-called $\mu\tau$ reflection symmetry \cite{mt1,mt2,mt3,mutaureflection_GL}, which exchanges a muon (tau) neutrino with a tau (muon) antineutrino in the charged lepton mass basis. 
If such a symmetry is imposed, the atmospheric mixing angle is predicted to be maximal, $\theta_{13}$ non-vanishing implies
a maximal Dirac phase $\delta$ and the Majorana phases vanish. The solar mixing angle $\theta_{12}$ and the reactor angle $\theta_{13}$ remain unconstrained.
\begin{figure}
\begin{center}
\includegraphics[width=18pc]{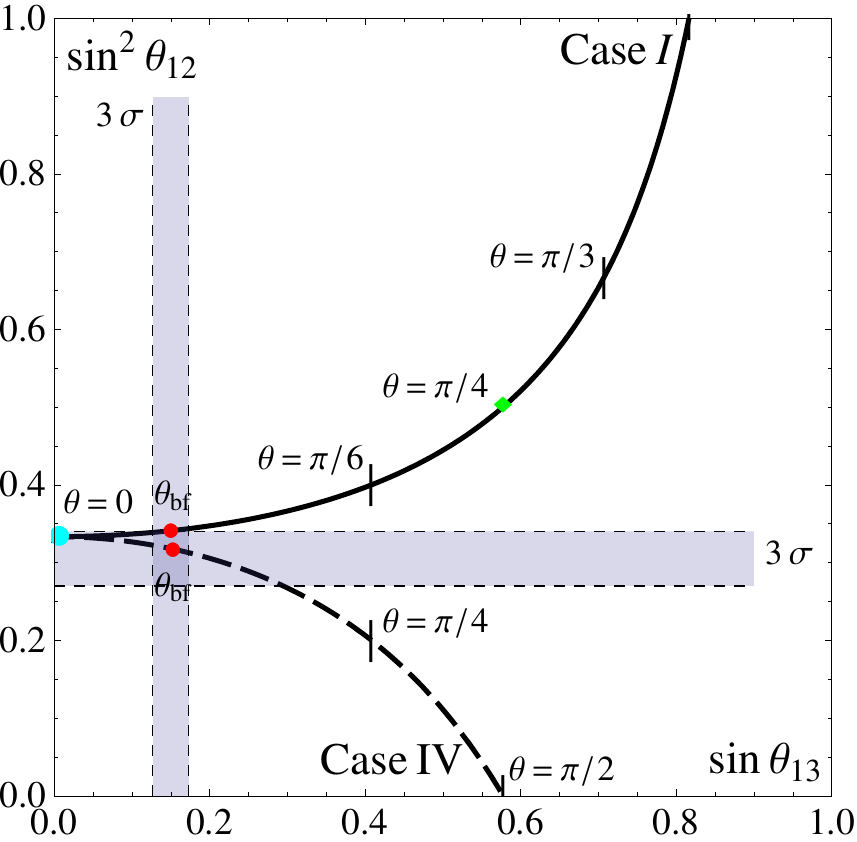}
\end{center}
\caption{\label{t4} Results for the mixing parameters $\sin \theta_{13}$, $\sin^2 \theta_{12}$  for Case I (straight line) and Case IV (dashed line), from ref. \cite{Feruglio:2012cw}. We mark the 
 value $\theta_{\mbox{bf}}$ of the parameter $\theta$ for which the $\chi^2$ functions have a global minimum with a red dot. $3 \, \sigma$ ranges for the mixing angles are also shown.}
\label{cp}
\end{figure}

\begin{table}[h!]
\caption{\label{ttt3} Lepton mixing angles and phases predicted from a theory invariant under $CP$ and the flavour group $\Delta(384)$, from ref. \cite{Hagedorn:2014wha}.}
\begin{center}
\begin{tabular}{ccccc}
\br
$\sin^2 \theta_{13}$&$\sin^2 \theta_{12}$&$\sin^2 \theta_{23}$&$\sin\delta$&$\sin\alpha=\sin\beta$\\
\br
$0.0220$&$0.318$&$0.579$&$+0.936$&$-1/\sqrt{2}$\\
\mr
$0.0220$&$0.318$&$0.421$&$-0.936$&$-1/\sqrt{2}$\\
\br
\end{tabular}
\end{center}
\end{table}

Flavour symmetries and generalized $CP$ transformations can be combined in a general formalism \cite{Feruglio:2012cw,GCPV1,GCPV2,CPGf_HD} constraining the lepton mixing matrix.
The starting point is a theory invariant under both $CP$ and discrete flavour transformations. After spontaneous symmetry breaking, by requiring appropriate residual symmetries in the neutrino and charged lepton sectors,
we can end up with a mixing matrix $U^0_{PMNS}$ completely determined up to one real parameter $\theta$ ranging from $0$ to $\pi$.
Mixing angles and phases, both Dirac and Majorana, are then predicted as a function of $\theta$, modulo the ambiguity related to the freedom of permuting rows and columns
and to the intrinsic parity of neutrinos. An exhaustive analysis of this formalism when the flavour group is $S_4$ has been presented in ref. \cite{Feruglio:2012cw}.
Two examples where the predicted mixing pattern have a maximal atmospheric mixing angle, a maximal Dirac phase and vanishing Majorana phases are shown in fig. \ref{cp}.
Recently several explicit models combining $CP$ and flavour symmetries have been proposed and several series of discrete groups have been investigated in combination with $CP$ \cite{Feruglio:2013hia,Ding:2013hpa,Ding:2013bpa,Ding:2013nsa,Chen:2014tpa,King:2014rwa,Hagedorn:2014wha,DiIura:2015kfa}. An interesting prediction obtained combining $CP$ with the flavour symmetry $\Delta(384)$ is shown in Table 2.

\section{Conclusion}
One of the weak points of the approach based on discrete symmetries is that they are mainly centered on the mixing matrix and there are no precise predictions for neutrino masses. 
In concrete models neutrino masses are typically only weakly constrained by the symmetry requirements that determine the LO mixing pattern.
Moreover there is no hint for such symmetries from quarks. Large hierarchies and small mixing angles do not seem to require discrete groups. 
As a consequence extensions to GUTs look quite involved. There are many existence proofs, but the discrete flavour group is typically badly broken in the quark sector.
Flavour symmetries represent certainly a useful tool, but no compelling and unique picture has emerged so far.
Despite many attempts to formulate a consistent and economic description of fermion masses and mixing angles, we are still far from a baseline model. 
Present data can be described within widely different frameworks, despite the constant, impressive, progress on the experimental side.
Simple schemes with a minimal amount of symmetry can well reproduce at the qualitative level the main features of the data in both quark and lepton sectors also in a GUT framework,
but no precision test can be set to assess their validity. Perhaps the right questions concerning the flavour problem and the specific aspect of neutrino masses and mixing angles have not yet been clearly identified.
We are still looking for some fundamental principle ruling the flavour sector, constantly supported in this effort by the continuous progress of the scientific community in eliminating unfounded ideas and improving the existing data.

\section*{Aknowledgements}
I warmly thank the organizers of Neutrino 2016 and of Now 2016 for giving me the possibility of presenting this talk. 
This work was supported in part by the MIUR-PRIN project 2010YJ2NYW, by the European Union network FP7 ITN INVISIBLES (Marie Curie Actions, PITN-GA-2011-289442) and by the Istituto Nazionale 
di Fisica Nucleare (INFN).

\end{document}